# Finite element modelling of shock - induced damages on ceramic hip prostheses

Juliana Uribe[1], Jérôme Hausselle[1], Jean Geringer[1]

[1]Center for Health Engineering, Biomechanics and Biomaterials Department

UMR CNRS 5146, IFR 143

Ecole Nationale Supérieure des Mines de Saint-Etienne,

158 cours Fauriel, 42023 Saint-Etienne, France

## ABSTRACT

The aim of this work was to simulate the behaviour of hip prostheses under mechanical shocks. When hip joint is replaced by prosthesis, during the swing phase of the leg, a microseparation between the prosthetic head and the cup could occur. Thus, when heel touches the ground, a shock up to nine times the body weight propagates through the leg. Two different sizes of femoral heads were studied: 28 and 32 mm diameter, made respectively in alumina and zirconia. The shock-induced stress was determined numerically using finite element analysis (FEA), Abaqus® software. The influence of inclination, force, material and microseparation was studied. In addition, an algorithm was developed from a probabilistic model, Todinov's approach, to predict lifetime of head and cup. Simulations showed maximum tensile stresses were reached on the cup's surfaces near to rim. The worst case was the cup-head mounted at 30°. All simulations and tests showed bulk zirconia had a greater resistance to shocks than bulk alumina. The lifetime was estimated with the probability of failure. This one could be bigger than 0.9 when a porosity greater than 0.7% vol. is present in the material. Simulating results showed good agreement with experimental results. The tests and simulations studied in this paper are promising for predicting the lifetime of ceramic prostheses.

# INTRODUCTION

Ceramics were investigated in orthopaedics since many years because of their good wear-resistance and loading capacity [1-4]. Ceramic-on-Ceramic, CoC, bearing surfaces reduce wear compared to Metal-on-Polymer, MoP, (The polymer is up today UHMWPE: Ultra High Molecular Weight PolyEthylene) or Ceramic-on-Polymer.

Nevertheless, retrieved heads and cups have sometimes shown wear stripes [5-8]. This pattern of wear has been shown to be related to a phenomenon of microseparation during the stance phase, which generates a contact force between the head and the edge of the cup [9]. This shock is believed to fracture the femoral cup, with a developed force up to 9 times the body weight.

Components for total hip replacement are shown in figure 1

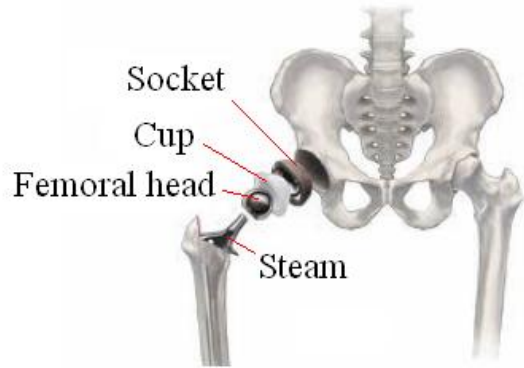

Figure 1 a) .Scheme hip articulation [10]

Clearly, the shock resistance issue requires a better understanding of the biomechanics of gait cycle. Hip joint forces have been estimated indirectly using inverse dynamics and analytical methods or directly measured with implanted transducers. In general, the force at the hip joint reaches an initial peak in early stance phase and a second peak in late stance phase. This pattern was described in the initial work of Paul et al [11, 12]. While the majority of the modelling work has been conducted on normal subjects, Paul et al. have also conducted a study on few subjects following total hip replacement [13]. However, there is a lack of studies describing how these forces generate stresses on the femoral components, and how it can lead to fracture of either one of the components.

Because of the increase of life expectancy, new materials or designs are needed, with a significantly improved lifetime and better mechanical properties. The goal of this study was to develop a model combining finite element analysis and crack growth simulation. The influence of several parameters was then assessed to determine the most significant parameter in relation with shock resistance.

## MATERIALS AND METHODS

### 1. Materials properties and designs

The degradation of two types of materials, alumina and zirconia, Table I, usually used for femoral heads and cups was simulated using FEA and compared to experimental tests results obtained using a shock machine [14], Figure 2 a). The orientation of heads and cups was set to vary between 30°, 60° and 45°, Figure 2 b). Three heads, 32mm made from alumina and three heads of 28 mm made from zirconia were tested. The cups corresponding to heads, in terms of diameter were investigated.

|  | Density (g/cm$^3$) | Young modulus (GPa) | Poisson's ratio |
|---|---|---|---|
| $Al_2O_3$ | 4.00 | 400 | 0.23 |
| $ZrO_2$ | 6.05 | 200 | 0.23 |

Table I. Physical and mechanical properties of alumina and zirconia.

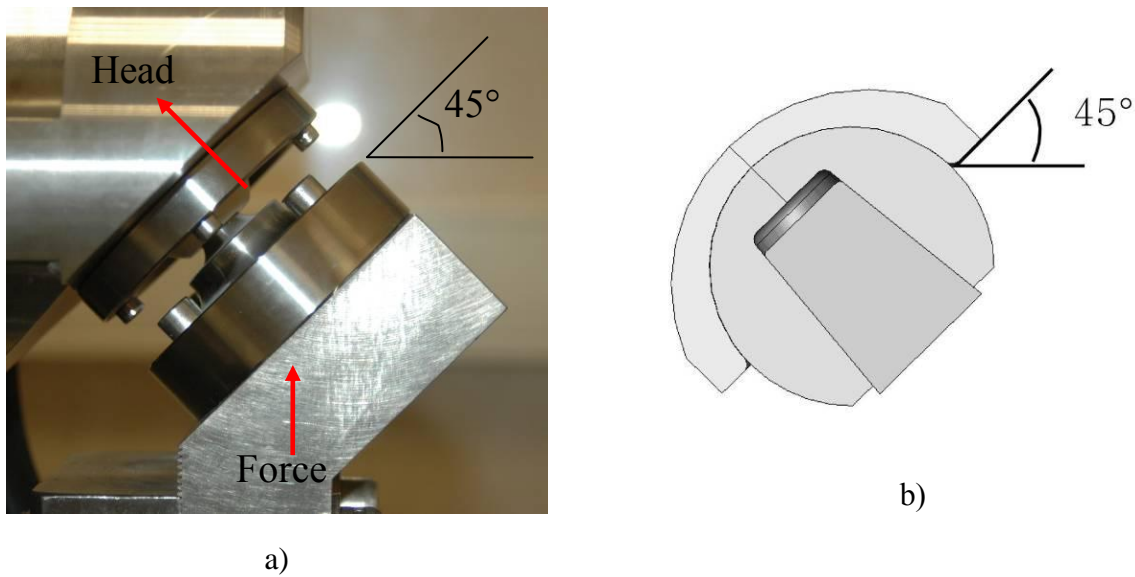

Figure 2. a) shock machine b) assembly head and cup at the standard anatomic position

### 2. Finite Element Modelling

The FE simulations were implemented using commercial software (ABAQUS 6.8, Dassault Systèmes ©, 2008)

• **Hypotheses**:

Materials were assumed to be perfectly elastic, so no plastic deformation was possible. The modelling was carried out in static conditions.
The shock-induced stress was considered constant. Finally, the friction between head and cup was considered negligible.

• **Loading and boundary conditions**

Friction coefficient for contact ceramic/ceramic is the weakest compared to other couples of materials. As the friction behaviour, in terms of Finite Elements Analysis (FEA), is not well known, the sliding movement was not considered in the contact. As the friction coefficient is weak, one might suggest that friction forces are negligible compared to contact forces, friction constant was zero

However a particular attention was paid on the elastic contact and the rebound, mechanical energy restored after a shock. An algorithm was developed using a bouncing ball made from zirconia. Head-taper fit and head cup were modelled with same contact algorithm penalty. This contact with the option "hard contact" allows modelling elastic contacts. This algorithm was developed using a ball made of zirconia which falls from a height of 10mm and impact. Coefficient of restitution was then calculated

$$e = \sqrt{\frac{h_{i+1}}{h_i}}$$

Where $h_i$ and $h_{i+1}$ are the height before and after bouncing repectively. This calculation was made for 8 bounces, and the coefficient of restitution was calculates as e=0.95+/-0.03, this value suggest the shock is purely elastic.
Even thought is possible to input a friction model in Abaqus, the model is unknown. That is why we said that there is not sliding in the contact. Moreover, the cup was constrained in all degree of freedoms to reproduce the shocks device.

In addition, symmetry conditions were imposed on the vertical axis. A load of 9kN was defined for the contact head-cup at the upper rim. Load was applied at the centre of the cone and increased linearly from zero to its maximum value at 11 ms which are the same conditions that experimental shock tests. Boundary conditions are shown in Figure 3.

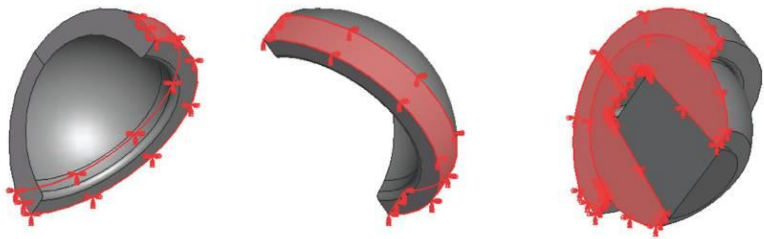

Figure 3.Boundary conditions: degrees of freedom and vertical symmetry.

• **Meshing**

Two different sizes were studied, 28 and 32mm in diameter. The first model consisted of a 28mm femoral head and a cup, both made of zirconia and a cone made of Ti-6Al-4V. Young's modulus and Poisson's ratio for zirconia were set to 200GPa and 0.23, respectively, see Table I. For titanium alloy, a Young's modulus of 110GPa and a Poisson's ratio of 0.32 were assigned. The mesh consisted of 4262 nodes in the head, 1927 in the cup and 169 for the cone. Second model consisted of a 32 diameter femoral head and a cup made in alumina. The Young's modulus for alumina was 400GPa and the Poisson's ratio 0.23. Head had 4799 nodes, cup 2534 and 145 for the cone. The shape of the rim was rounded with a radius of 1mm. Figure 4a shows the three dimensional assembly.

Both FEA models were three-dimensional and meshed using tetrahedral solid elements, ideal for meshing complexes geometries. In order to obtain more accurate results, the mesh in the hypothetical contact zones between head and cup were refined, especially at the rim of the cup. Meshing of head and cup are shown in figure 4.

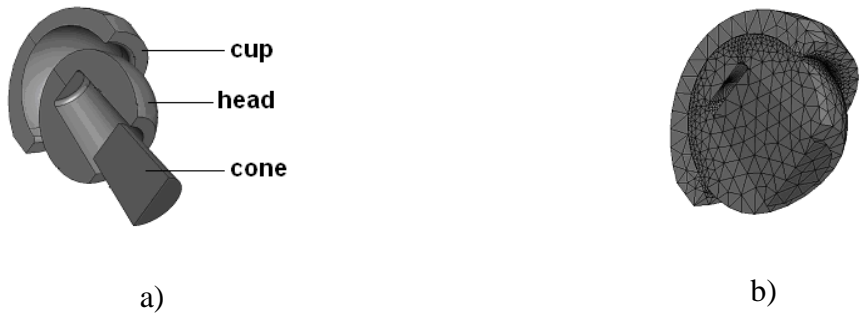

a)                                                     b)

Figure 4. a) Assembly of the cup, head and cone for the finite element simulation
b) Refined mesh of the assembly

Several parameters were changed in order to determinate their influence on the stress. These parameters are summarized in table II.

| Parameter | Material | Force | Microseparation (mm) | Inclination |
|---|---|---|---|---|
| **Force** | zirconia | 2-9 kN | 1.3 | 45° |
| **Microseparation** | zirconia | 9 kN | 0, 0.7, 1.0, 1.3, 1.6, 1.9 | 45° |
| **Inclination** | zirconia | 9 kN | 1.3 | 30°, 45°, 60° |
| **Material** | zirconia alumina | 9 kN | 1.3 | 45° |

Table II. Influence of force, inclination, material and microseparation on the cup's stresses

- **Treatment/post processing with Matlab®**

An algorithm implemented in Matlab (Matlab 7.7 The MathWorks Inc®) was used to interpolate stresses on the nodes obtained with ABAQUS®. This step is necessary for increasing the number of nodes in order to predict the structure behaviour. For each element, stresses in the nodes were interpolated into a cube (Fig 5a and b). Then, an algorithm allowed keeping only the points into the finite element (Fig 5c).

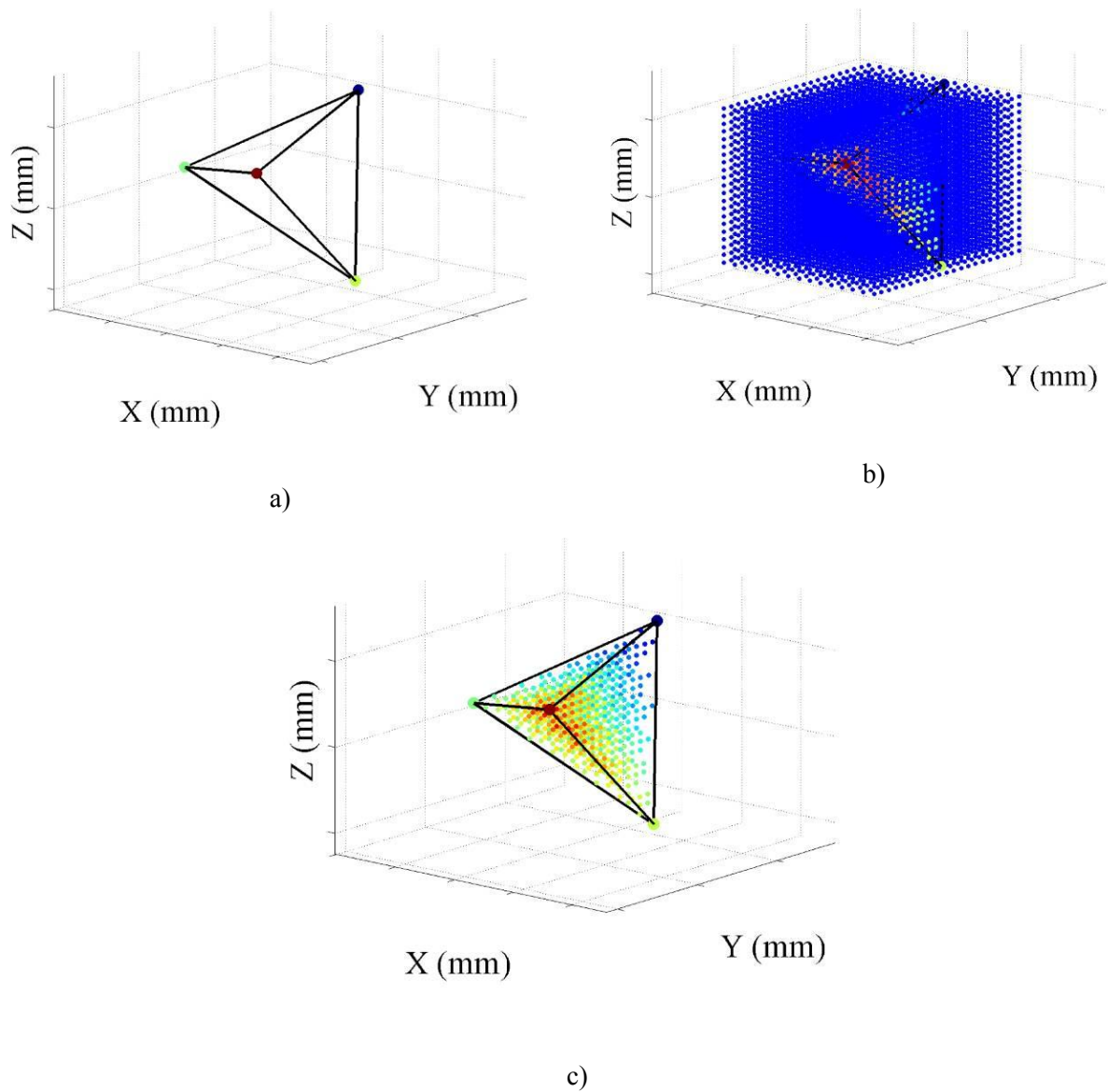

a)

b)

c)

Figure 5.Interpolation 3D of stresses on the cup a) A finite element: colour represents stress value; from red for the minimum to blue for the maximum stress b) interpolation cube including the element, nodes outside of the cube are plotted blue c) remaining nodes after deleting nodes out of the cube.

Thus, stresses on the cup were calculated in about 3 millions of points instead of a few thousands. Number of points was multiplied by 1000.

### 3. Algorithm

• General considerations

Stress intensity factor, $K_I$, is defined as:

$$K_I = Y\sigma\sqrt{\pi a} \qquad (1)$$

Where:
Y is a constant which takes into account the geometry and location of the flaw, considered as 1.
σ is the tensile stress
a is the depth for a crack or radius for a porosity.

Only tensile stresses were considered. Stress is supposed to be maximum and constant during the 11ms of the simulation. Therefore, $K_{max} = K_I$, we have then:

$$V = AK_I^m \qquad (2)$$

An yttria-doped zirconia has m=20, and from experimental curves [15], the constant A=1.2*10$^{-18}$ was calculated. Under cyclic solicitations, the ratio between $K_{I0}$ and $K_{IC}$ is 0.48 [16]. Since $K_{IC}$ is reported to be $5.5 MPa\sqrt{m}$ [17], $K_{I0} = 2.7 MPa\sqrt{m}$ was fixed for simulating cyclic solicitations.

For alumina, the same ratio between $K_{I0}$ and $K_{IC}$ was taken. Hence, $K_{I0} = 2.1 MPa\sqrt{m}$ was considered. Coefficients A and m were taken from a V vs. $K_I$ curve for alumina under cyclic solicitations [16]. Coefficients used for alumina and zirconia are summarized in Table III. Curves used in simulations are shown in Figure 6.

|  | A | m | $K_{IC}$ ($MPa\sqrt{m}$) | $K_{I0}$ ($MPa\sqrt{m}$) |
|---|---|---|---|---|
| Al$_2$O$_3$ | 1.3x10$^{-23}$ | 40.0 | 4.2 | 2.1 |
| ZrO$_2$ | 1.2x10$^{-18}$ | 20.0 | 5.2 | 2.7 |

Table III. Parameters for V-$K_I$ curves

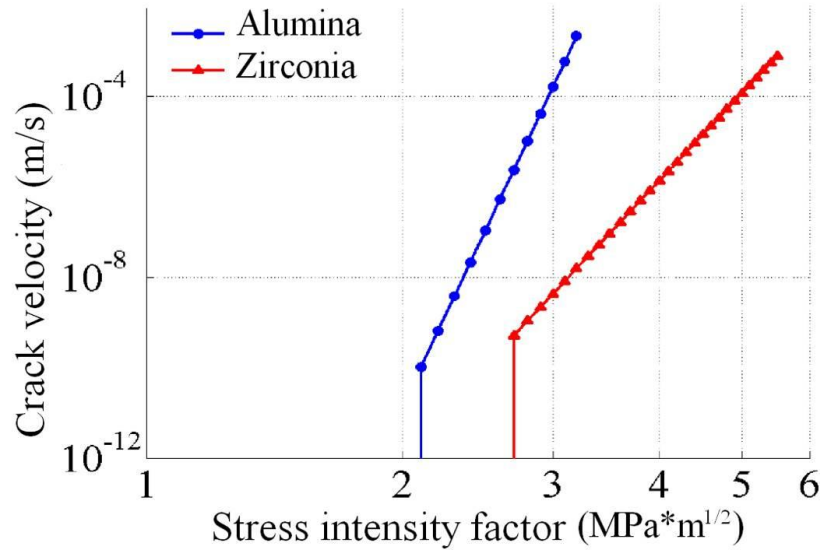

Figure 6. Curves V vs. $K_I$ used for alumina and zirconia crack growth simulations.

• **Algorithm**

The configuration is considered unstable and the associated flaw will growth while $K_I$ is higher than $K_{I0}$. If $K_I$ is smaller than $K_{I0}$, then tensile stress is too small to involve the flaw growth, so the configuration is considered stable. Thereafter, calculations were launched only for the unstable configurations Ne, Figure 7.
For each shock N and every unstable configuration Ne, the stress intensity factor is calculated, the crack velocity is deducted and the new flaw size is determined. With this new flaw size, the stress intensity factor is recalculated and compared to $K_{IC}$. If $K_I$ is higher than $K_{IC}$, then the critical flaw size is reached and the cup is considered fractured and the fractures counter nF increases by 1, Figure 8. If $K_I$ is smaller than $K_{IC}$, then, the configuration runs again. Therefore, the next configuration is considered. Finally, when all configurations are tested, the number of shocks increases and the calculation begins again for all the unstable configurations.

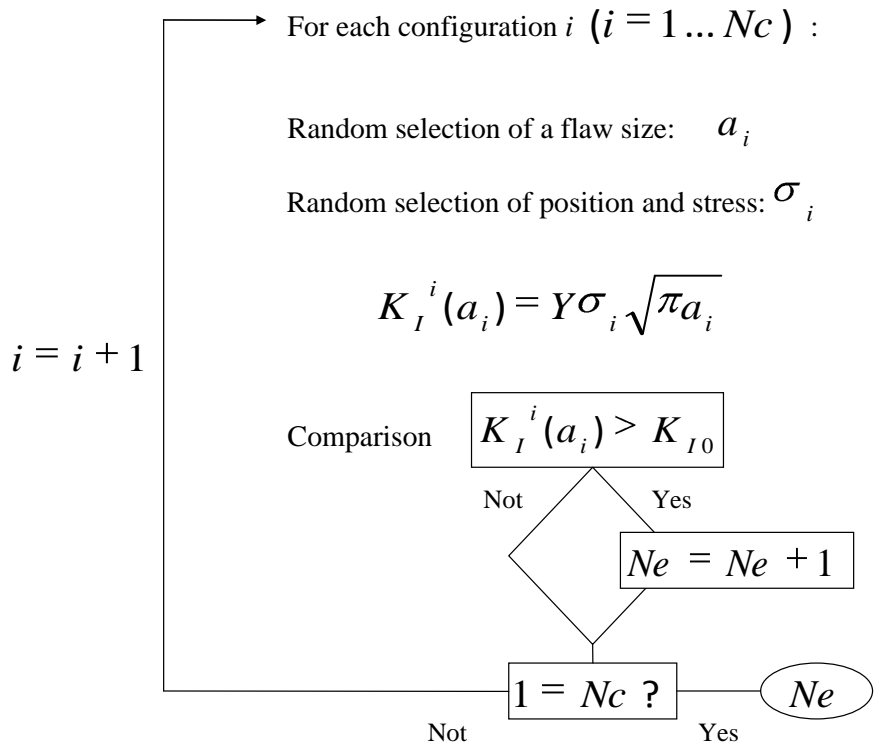

Figure 7. Algorithm for determining the number of unstable configurations.

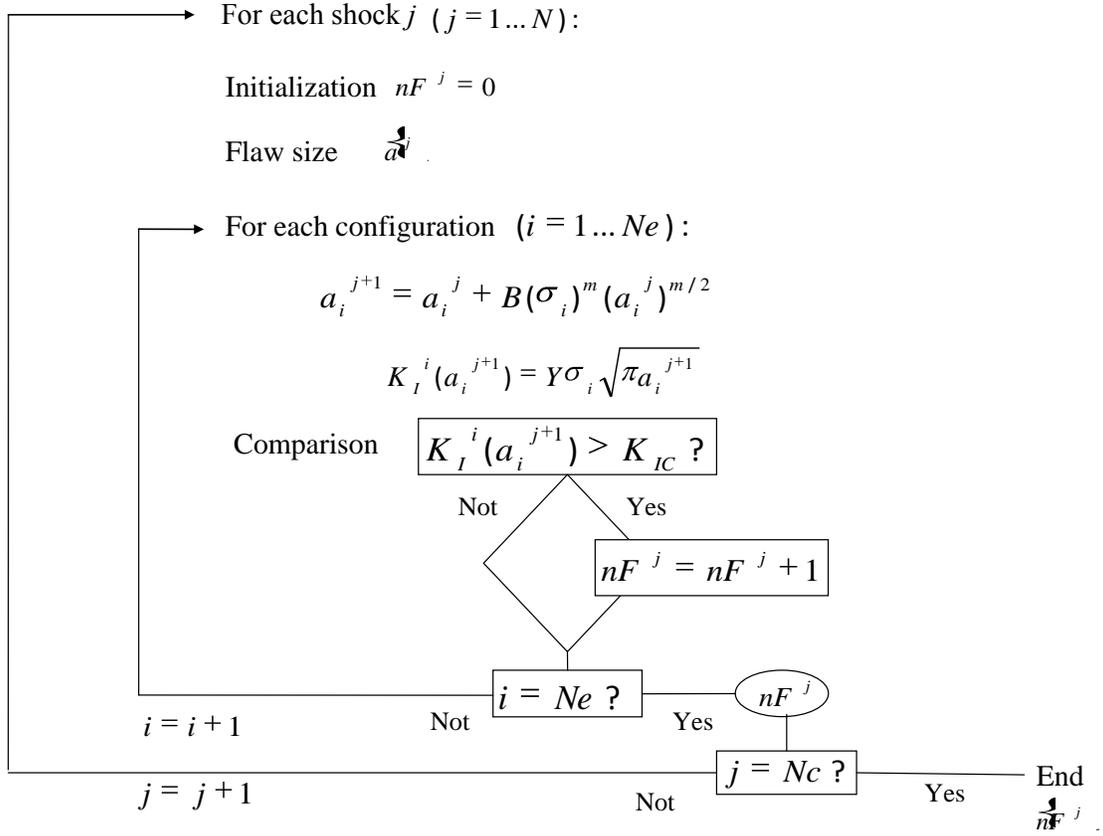

Figure 8. Algorithm for crack growth simulation

• **Verifying the systems determinism**

Since the algorithm consists of iterating many times a markedly nonlinear function, it is necessary to verify the determinism of calculations. That is why a relation between a flaw size before and after the shock j+1 is established:

$$a_{j+1} = a_j + v_{j+1} \Delta t \qquad (3)$$

From the Eq. 1 and 2:

$$v_{j+1} = A Y^m \sigma^m \pi^{m/2} a_j^{m/2} \qquad (4)$$

Hence:

$$a_{j+1} = a_j + B \sigma^m a_j^{m/2} \quad ; \text{Where } B = A Y^m \pi^{m/2} \Delta t \qquad (5)$$

The parameters A and m included the zirconia and alumina values given by the Table III. The aim was to verify that the number of shocks for cracking is a monotone function of A, m and initial flaw size. Variation in the number of shocks for cracking would mean that the equations describe a chaotic system not predictable in time [18]. After testing the algorithm, it is worth noting that the equations showed to describe a deterministic system useful in estimating the prostheses lifetime.

• **Optimum of configurations number, Nc**

Given that size and location of the flaws are randomly determined, test simulations were investigated in order to assess the optimal number of configurations Nc to use. Those simulations were performed for an assembly of head-cup made from zirconia and inclined of 45°, with an applied force of 9kN, a microseparation of 1.3mm and a flaw initial average size of 40μm. Hundreds of simulations were made for every number of configurations. The growth percentage, the standard deviation and the calculation time are shown in Figure 9 for Nc=$10^3$, $5 \times 10^3$, $10^4$, $5 \times 10^4$, $10^5$, $2 \times 10^5$ and $5 \times 10^5$.

The configurations number would be a compromise between the calculation time vs. the growth percentage and the standard deviation. According to the results from Figure 9, one might suggest for considering Nc=$10^5$ should be the best compromise in order to obtain a good repeatability of results with a moderate calculation time. Average calculation time for the $10^5$ configurations after shocking is 30s, which means 0.3ms per configuration per shock. Since the average growth percentage is about 0.2%, there will be 200 unstable configurations, which means a calculation time of 60ms per shock. Thereby, the complete simulation of $10^5$ shocks might require a calculation time of one hour and 35 minutes, if there is not pre-existing cracks.

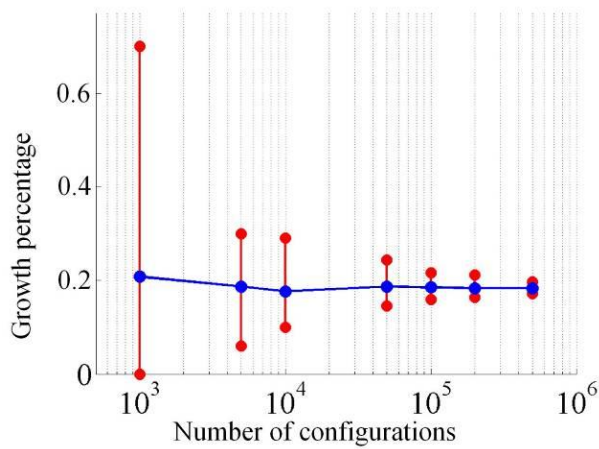
a)

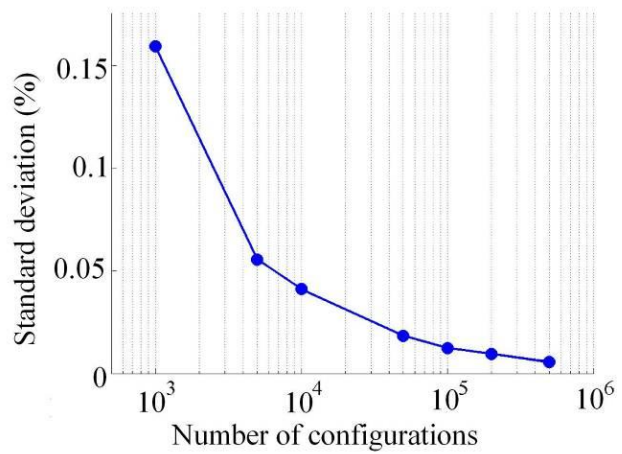
b)

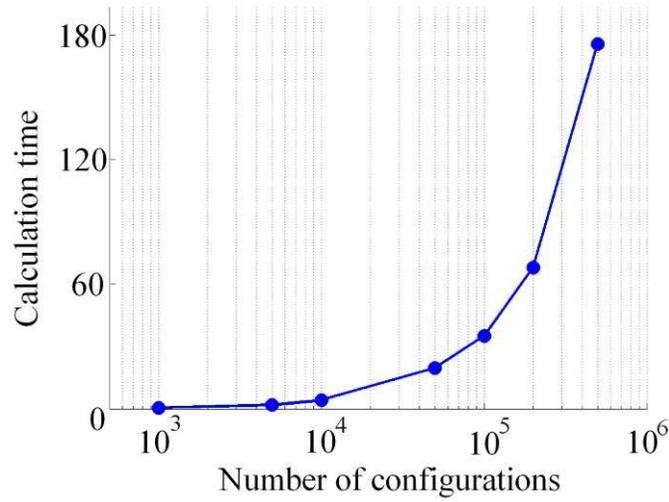

c)

Figure 9. Determination of optimal number of configurations a) Average percentage of growth as function of the simulated number of configurations. Red dots represent minimum and maximum of percentage for every number of configurations b) Standard deviation of average percentage of growth c) Average calculation time for detecting unstable flaws (according to the number of configuration)

## 4. Probability of failure

The previous algorithm, described part 2, allows getting the $N_c$, number of tested configurations, and the $N_F$, number of failures. Thus the F, the individual conditional probability, should be defined by

$$F = N_F/N_C \qquad (6)$$

Moreover, thanks to probabilistic model developed by Todinov [19][20], the probability evolution, P, might be calculated.

By making the hypothesis that there is not flaws coalescence and they evolve independently, the probability of failure, P, for an element under a constant loading σ is:

$$P = 1 - \exp[-\lambda V F] \qquad (7)$$

Where λ is the density of flaws (the location of flaws is in accordance to a homogeneous Poisson distribution). V is the volume and F is the individual conditional probability to involve a failure from a unique flaw [19]. This probability is conditional because it represents the probability of failure initiated by flaws, with the hypothesis that this flaw is into the volume V. Flaws density λ is calculated with the material porosity $p$, and the flaw average volume $V_\alpha$, both supposed spherical:

$$\lambda = \frac{p}{(1-p)V_\alpha} \qquad (8)$$

Assuming a porosity of the bioceramics from 0.01% to 1.00% and flaws average size between 1 and 100µm, λ is between 0.2 and 2.5x10$^6$, Figure 10.

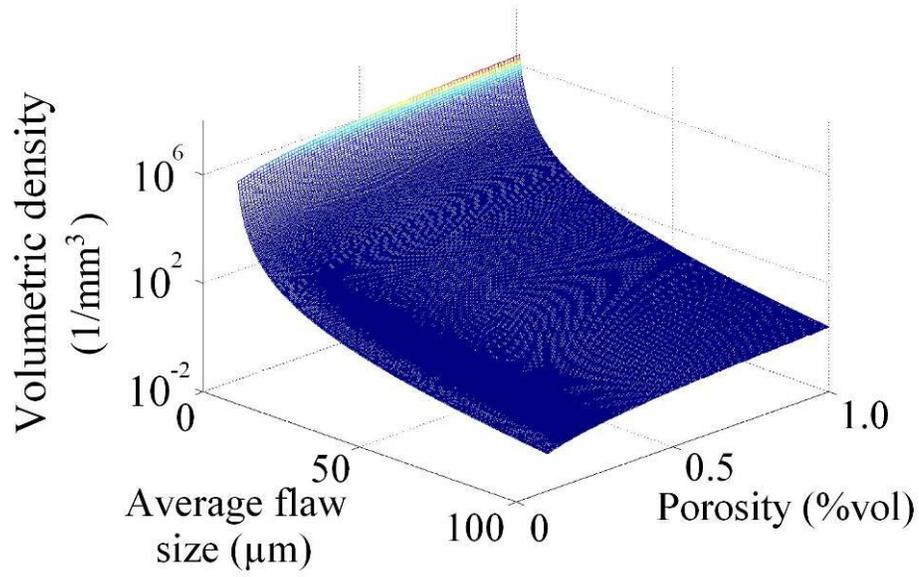

Figure 10.Determination of flaws density according to porosity and average flaw size

# RESULTS

1. **Stresses distribution using Finite Element Analysis (FEA): ABAQUS ®**

Various configurations will be considered in this part. One factor changes and the others are constant: force, microseparation, angle and materials. In this part, the flaws increasing will not be mentioned. One focuses our attention on the influence on the experimental parameters to better understand the influence of each one.

• Force, zirconia 28 mm

In order to validate the crack growth model from experimental results, some modellings were investigated for a cup tilted at 45°, with former microseparation from 0.0 to 1.3mm and a force between 2 and 9kN. First, a contact between the head and the upper rim of the cup takes place. Then, the head impacts the upper rim of the cup, rebounds and impacts the lower rim at the end of simulation, the duration of this movement is 11ms.
Stresses on the cup are shown for the directions theta and phi, figure 11.

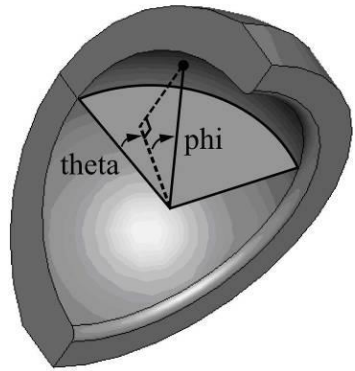

Fig 11. Angles theta and phi on the head

On the cup, the maximum stresses were always located near the rim, Figure 12 a). The applied force did not change the stresses location but only its magnitudes. Moreover, for 2kN, stresses are located on the upper rim of the cup. In fact, the shock is not strong enough to involve the head rebound and impact the lower rim.
For higher forces, stresses are also observed on the lower rim of the cup. However, stresses on the upper rim are slightly greater than those on the lower rim, Figure 12 b).

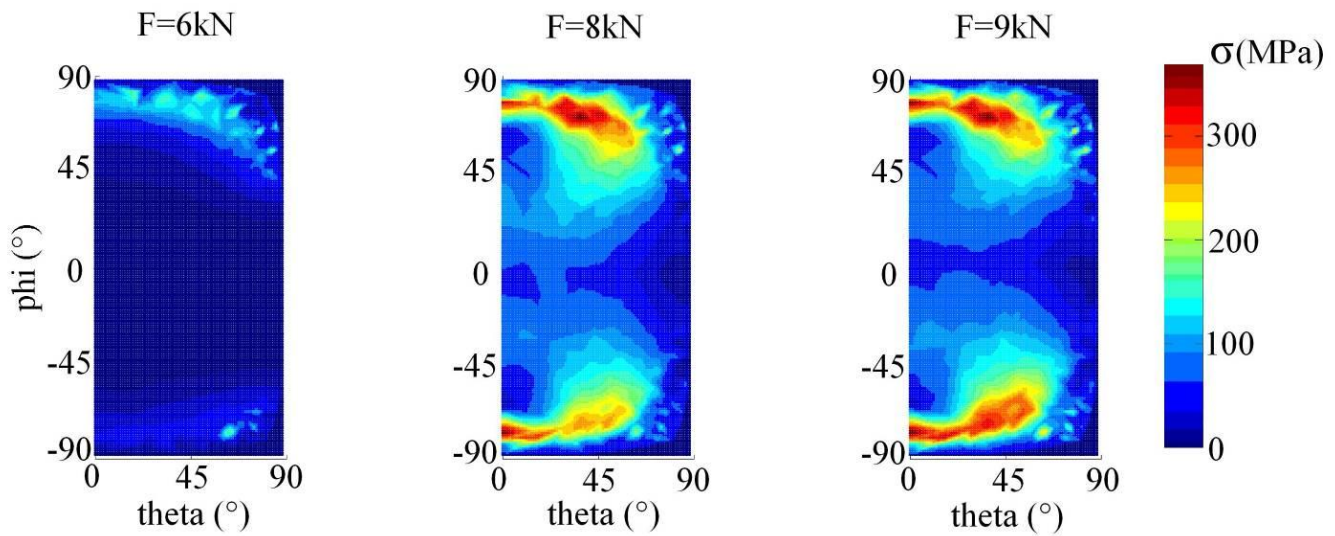

a)

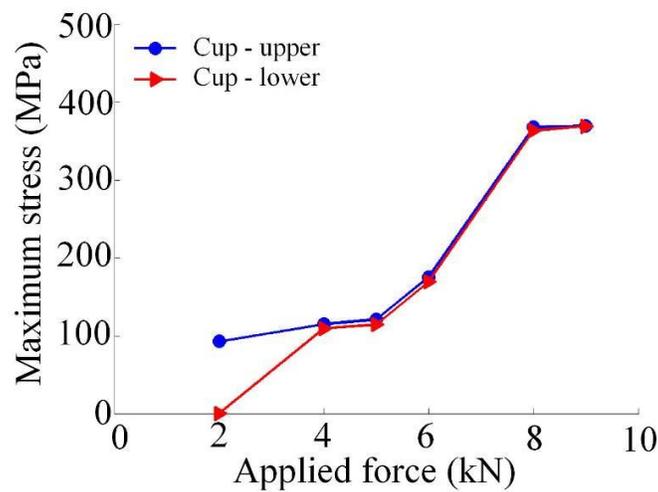

b)

Figure 12.Influence of applied force on stresses on the cup, microseparation of 1.3 mm and angle of 45° a) Location of tensile stresses on the cup under a force of 6, 8 and 9 kN b) Maximum stresses as function of the applied force.

This phenomenon, probably related to rebounds at the lower and the upper rims, is likely to appear in the experimental tests since two wear stripes were located on the head. These results corroborate exactly some explants studies [5, 21, 22], after each impact the stresses are not constant, thus the maximum stresses are time dependent.

• Microseparation, zirconia 28mm

The influence of the microseparation was investigated with a cup position of 45° and a force of 9 kN. Three vertical displacements were 0, 0.5, 0.75, 1.0, 1.25 and 1.5mm, which correspond respectively to these microseparations: 0, 0.7, 1.0, 1.3, 1.6 and 1.6mm. As mentioned above, stresses are located on the upper rim and the lower rim of the cup, Figure 13. The lower microseparation multiplied the stress by a factor of 100 on the cup's surface, compared with the case without microsepration.

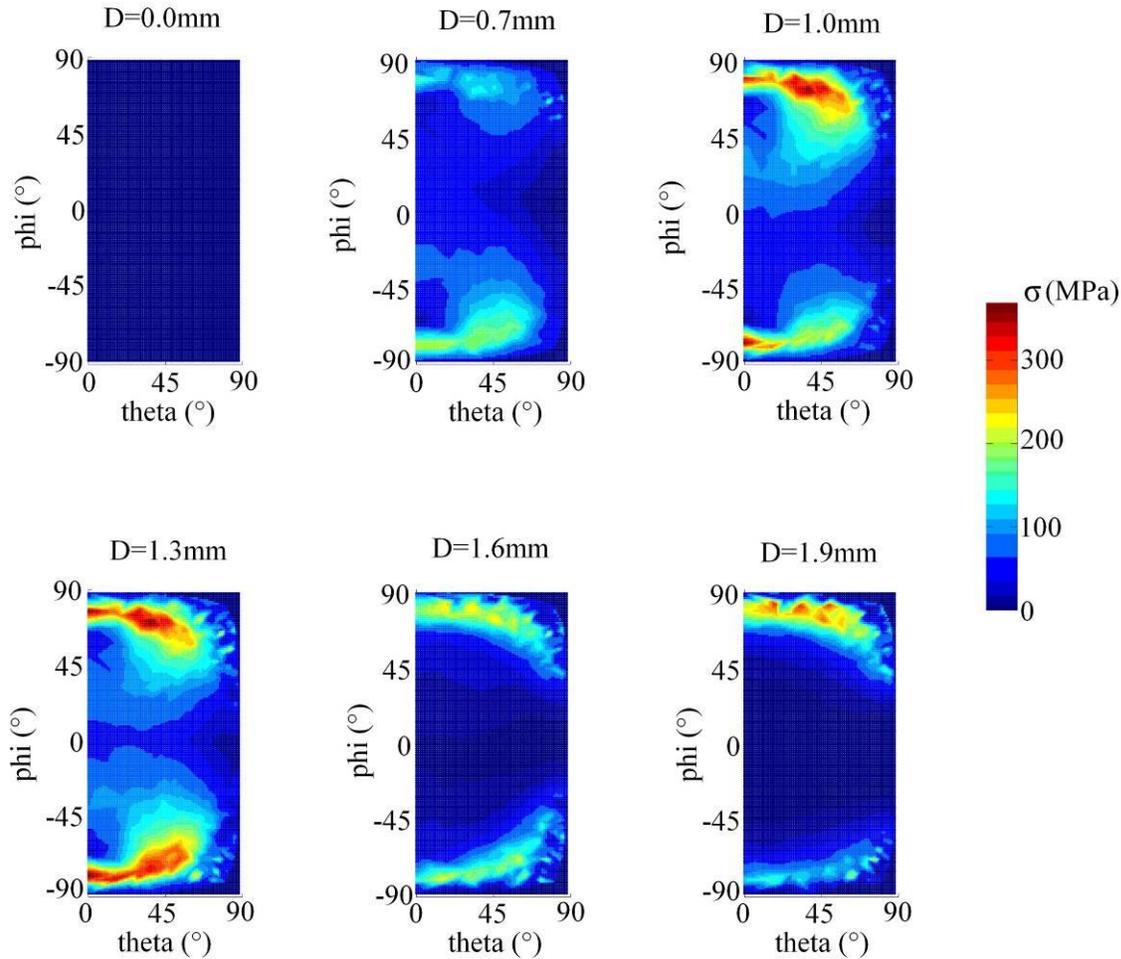

Figure13. Influence of microseparation on the surface stresses of the cup; tensile stresses on the cup for 0, 0.7, 1.0, 1.3, 1.6 and 1.9mm.

Tensile stress of 400MPa in Fig 13 was obtained after first impact. Stresses are shown for the upper and lower rim of the cup. The femoral head impacts the upper rim of the cup and then the lower one and so on until the end in t=11ms. This phenomenon has not been proved *in vivo* but it might exist when testing bearing surfaces in shock device since two wear stripes are observed on the head (upper and lower). Maximum stresses are located near the upper rim on the cup. The worst case leading to biggest stresses is a microseparation between 1.0 and 1.5mm. This threshold of microseparation for maximum stress is valid for bearing surfaces of diameter 28mm. For femoral heads with a different geometry, the value of microseparation leading the maximum stresses could vary. Moreover, in the finite element modelling, the cup is constrained in all degrees of freedom while in the shock device the cup has one degree of freedom in the axis perpendicular to the applied force. Some investigations are in progress in order to determine the elastic response of the cup in the shock device. Thus, an elastic and damping coefficients could be input to the finite element model, modifying the boundary conditions and leading to different values of stresses on the cup.

• Inclination, zirconia 28mm

The influence of the cup angle, the inclination, was studied for 30°, 45° and 60, Figure 14a). For each angle, the load was of 9kN. The modellings were investigated for several microseparation values: 0, 0.7, 1.0, 1.3, 1.6mm. The location of the maximum stress, on the cup surface, does not depend on the inclination of the cup, however the maximum value does (figure 14 b).

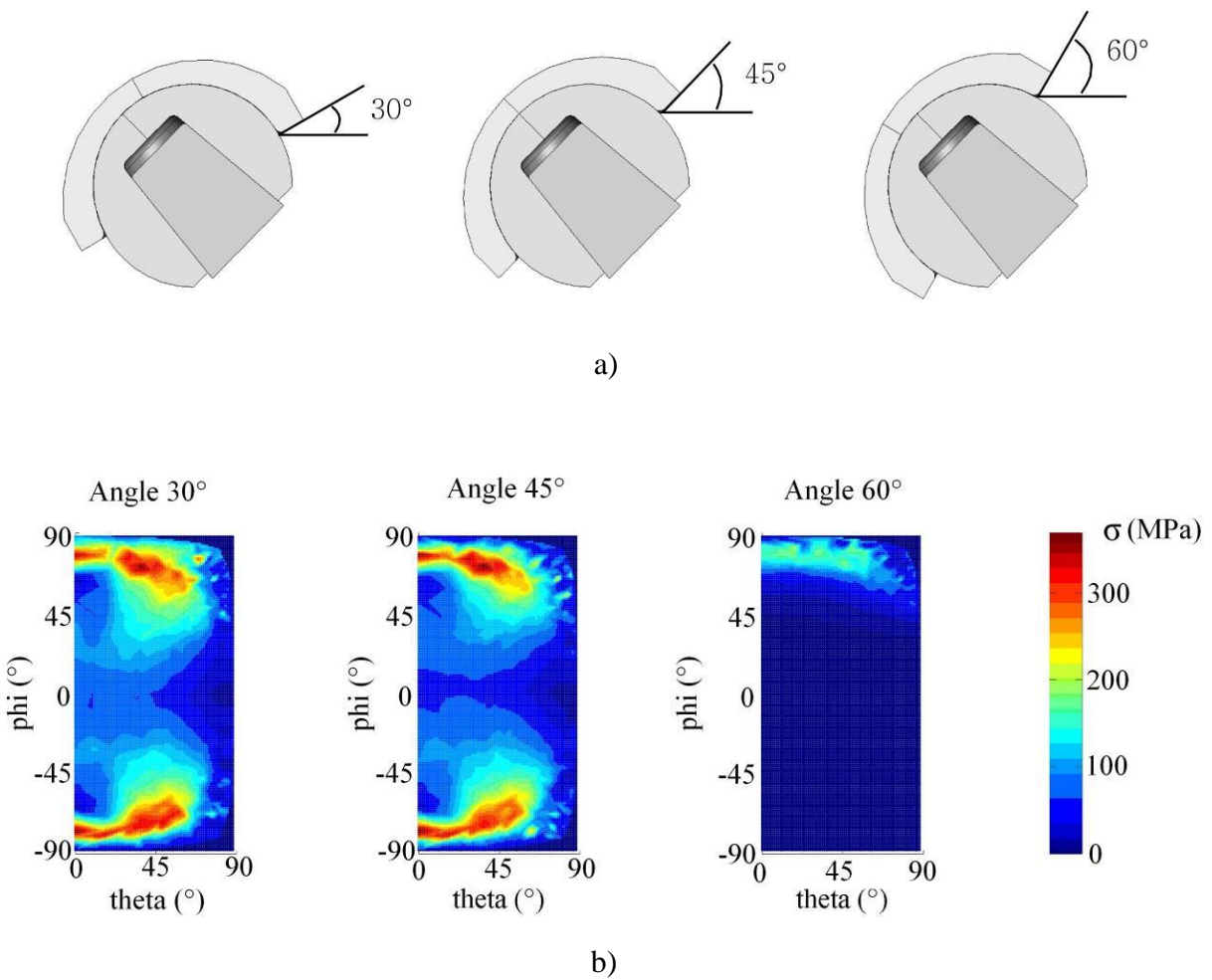

Figure 14. Influence of inclination on the surface stresses of the cup. a) Tested inclinations. b) Stresses on the surface of the cup for each inclination and a microseparation of 1.3mm

• Material, zirconia 28mm and alumina 28mm

An alumina-alumina prosthesis of 28mm was considered, mounted at 45°, with a force of 9kN. Stresses on the cup were 30% higher than stresses on the zirconia one. Results are shown on Figure 15

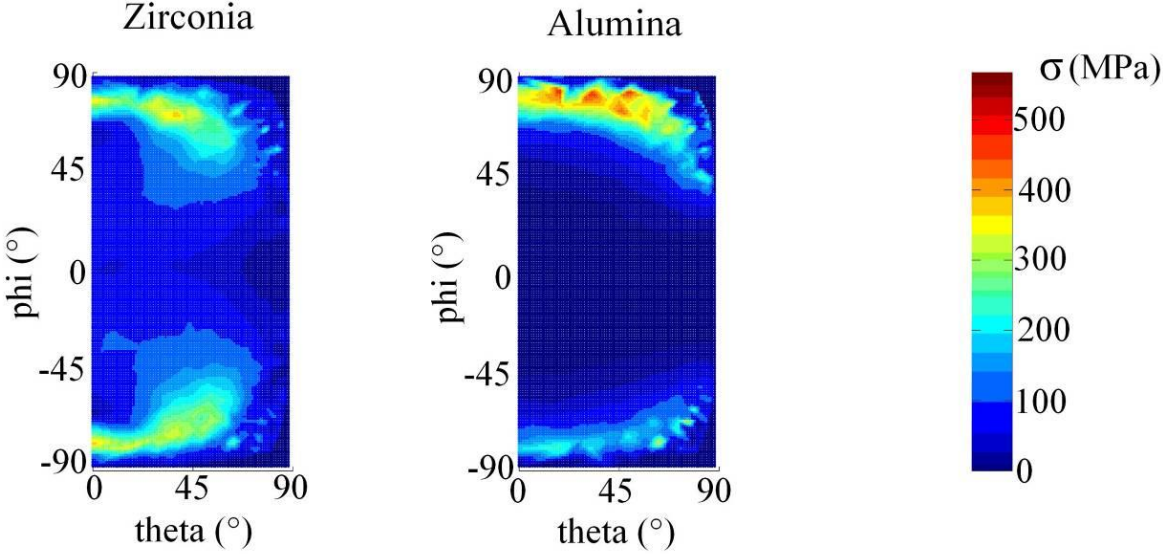

Figure 15. Influence of material on the surface stresses on the cup, with a load of 9kN, inclination of 45° and microseparation of 1.3mm

## 2. Algorithm of crack growth using Matlab ®

In this part, the Todinov's approach is investigated for predicting the crack growth according to a flaws distribution. Positions of the flaws are shown on figure 16. Both kinds of coordinates are considered: spherical coordinates (phi-theta) at the left and cartesian coordinates (x-y) at the right. Colored dots represent the depth of the flaw inside the cup, according to the scale at the right of the graphs.

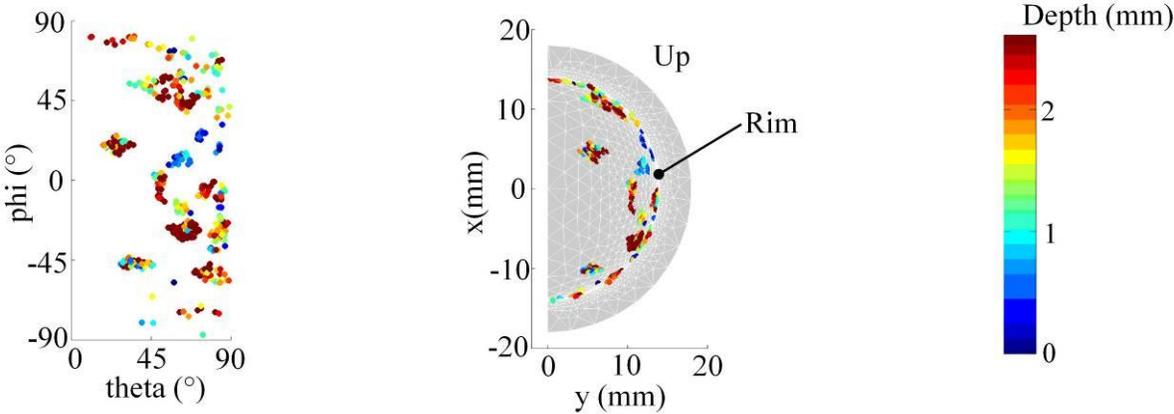

a)

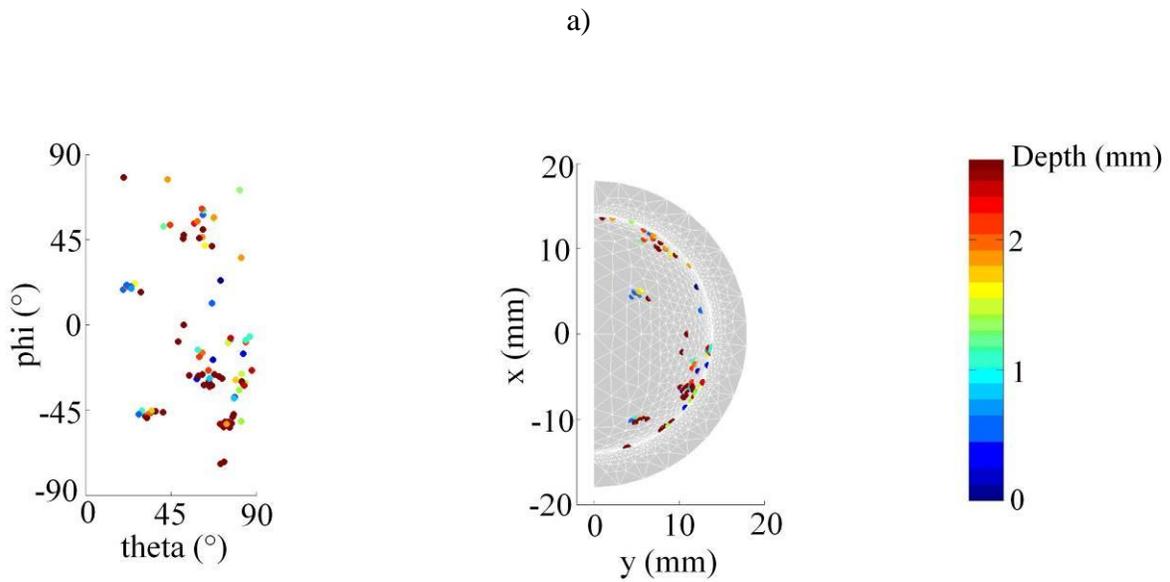

b)

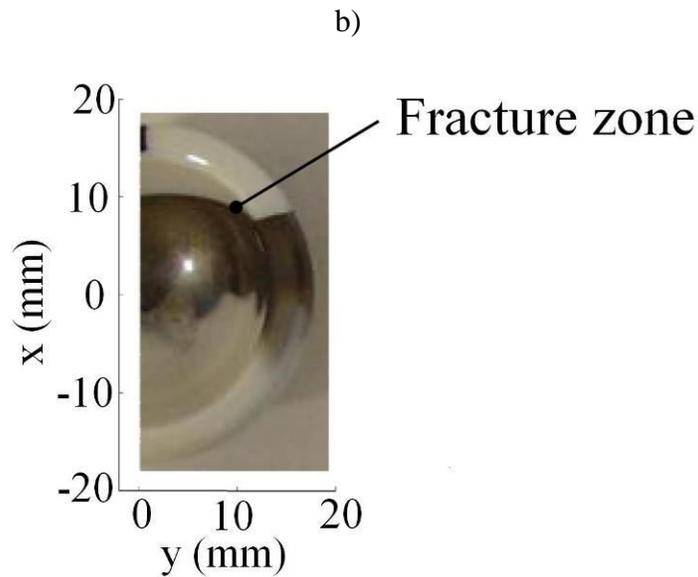

c)

Figure 16. Location of simulated flaws, cup inclined at 45°, load of 9kN and microseparation of 1.3mm. a) Location of critical flaws for a flaw size between 19 and 35 µm. b) Location of critical flaws for a flaw size between 24 and 27 µm. c) Crack surfaces observed experimentally on an alumina cup.

There was no critical flaw deeper than 2mm. Moreover, these critical flaws seemed to be distributed on almost all the surface of the cup, Figure 15a. However, when filtering for having flaws between 24 and 27µm, the distribution is limited to both zones: upper and lower surface, Figure 16b.

The Figure 16c highlights the fracture on a tested cup is in the same zone than the ones showed from the modelling, Figures 16a and 16b.

**Probability of failure**

**• Influence of porosity on failure probability**

As mentioned before, material flaws density directly depends on the material porosity (Eq 8). Calculations were made with several porosities, ranging from 0.1% to 1.0% vol, Figure 17. For a porosity higher than 0.7%, the less critical flaw leads to a failure probability higher than 0.9.

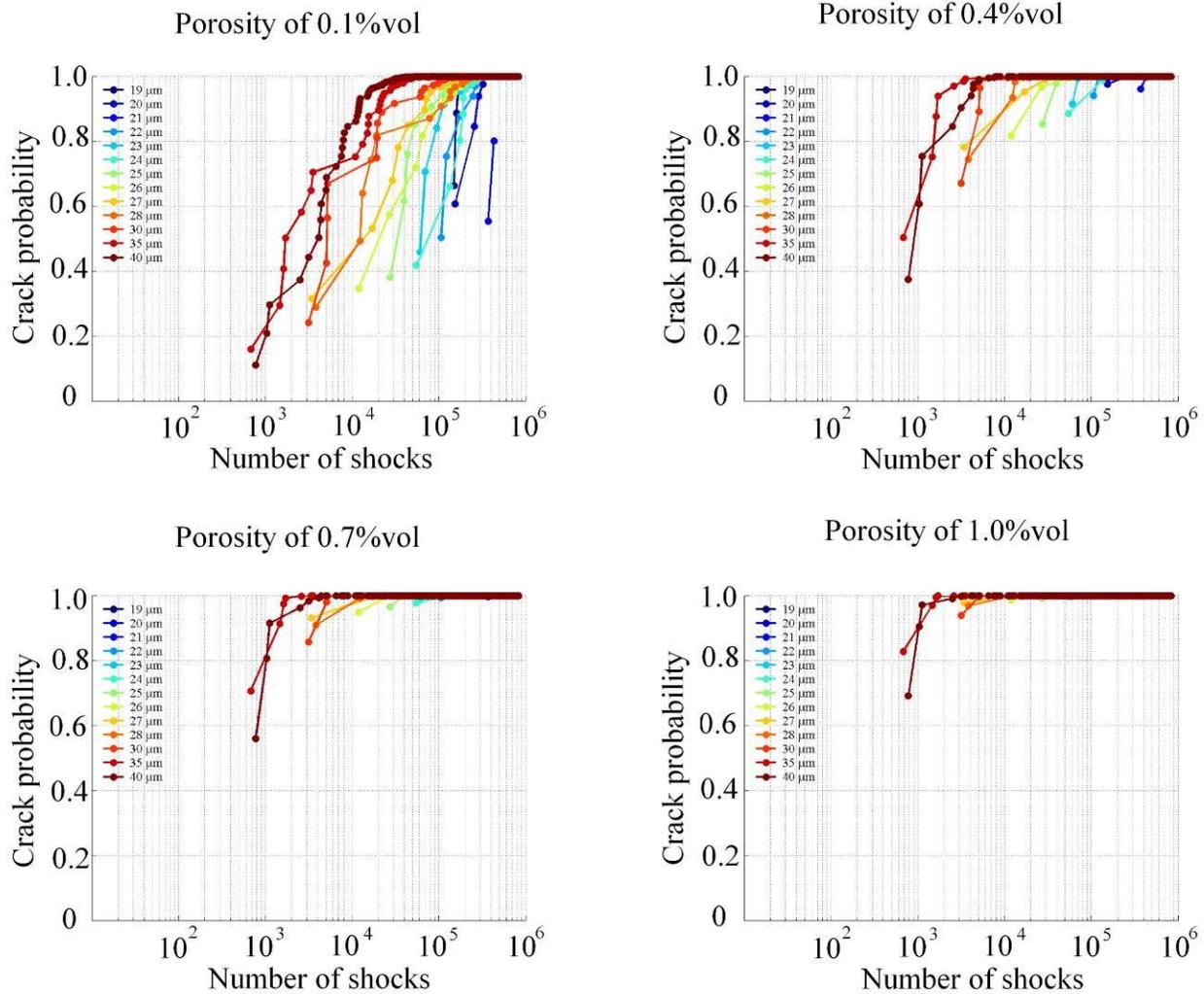

Figure 17.Evolution of crack probabilities vs. the number of shocks, according to flaw size (from 19 to 40 μm) and porosity (from 0.1 to 1 %); all following parameters are constant, inclination angle of 45°, load of 9kN and microseparation of 1.3mm.

**• Influence of inclination on failure probability (zirconia 28 mm)**

Flaw growth model was performed to study the influence of the inclination on the lifetime estimation. The inclinations of 30°, 45° and 60° were considered, with a constant applied force of 9kN and a microseparation of 1.3 mm. It is worth noting that the flaws sizes are in the range from 20 to 40 µm for the inclinations of 30° and 45° (Figure 18a and b). For these both inclinations, the evolution of failure probability during shocks is roughly identical. The probability of failure as a function of the number of shocks is showed in Figure 18c.

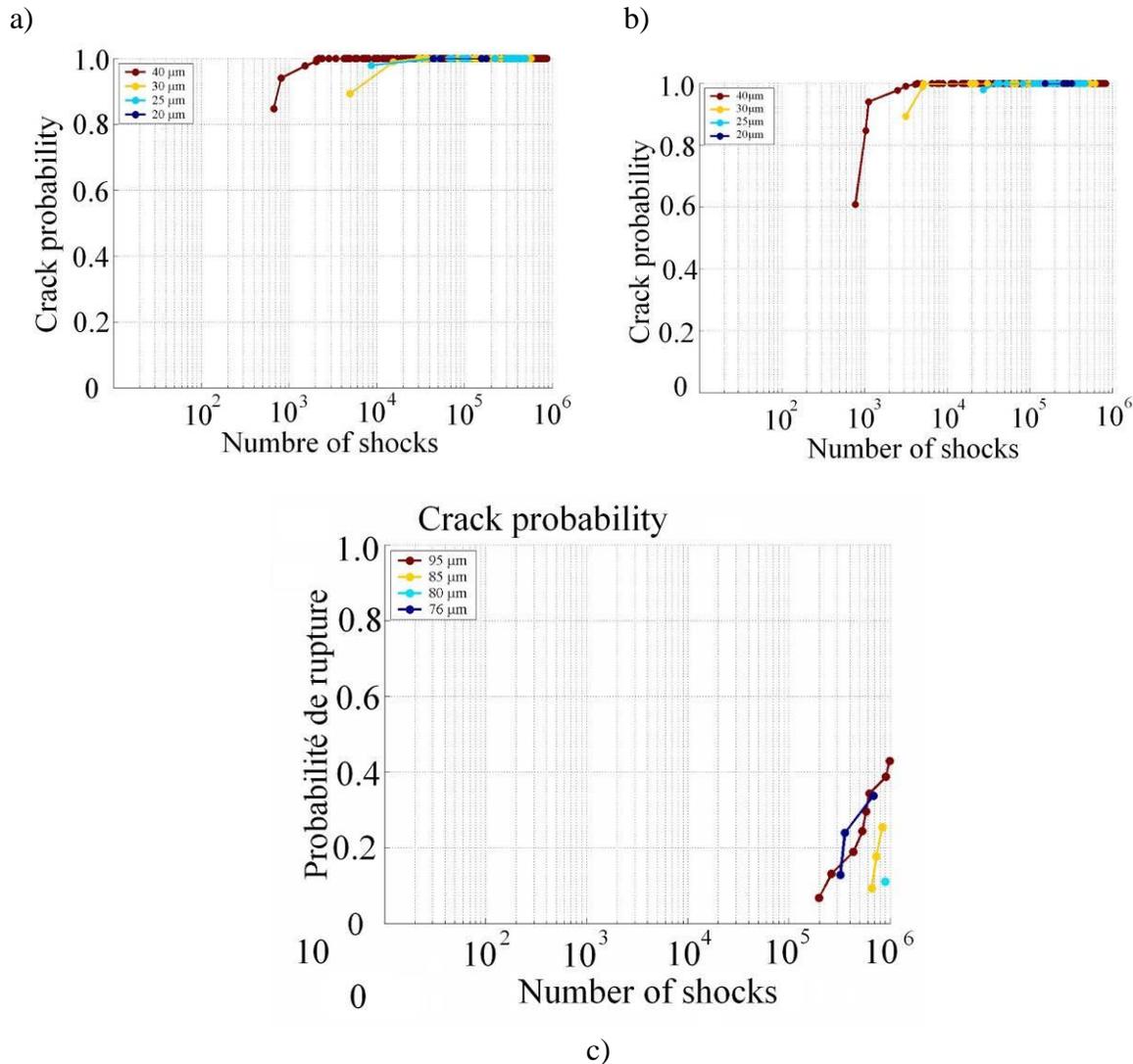

Figure 18. Probability of failure as a function of the number of shocks for a) a cup inclinated at 30° and initial flaw sizes of 20, 25, 30 and 35µm b) a cup inclinated at 45° and initial flaw sizes of 20, 25, 30 and 35µm c) a cup at 60° and initial flaw sizes of 76, 80, 85 and 95µm

Furthermore, the small flaws size makes the probabilities growth quickly toward the unity, since the density of flaws is high. For an angle 60°, number of shocks before any flaw reaches the critical size, is higher than those for 30° or 45°. In addition, flaw sizes likely to growth have to be higher than the ones for 30 and 40

• **Influence of material on failure probability**

The modelling parameters are as following: zirconia-zirconia and alumina-alumina, head diameter of 28mm, inclination of 45°, force of 9kN and microseparation of 1.3mm. The minimum flaws sizes likely to growth in alumina are smaller than those for zirconia: 4µm and 19µm respectively, Figure 19.

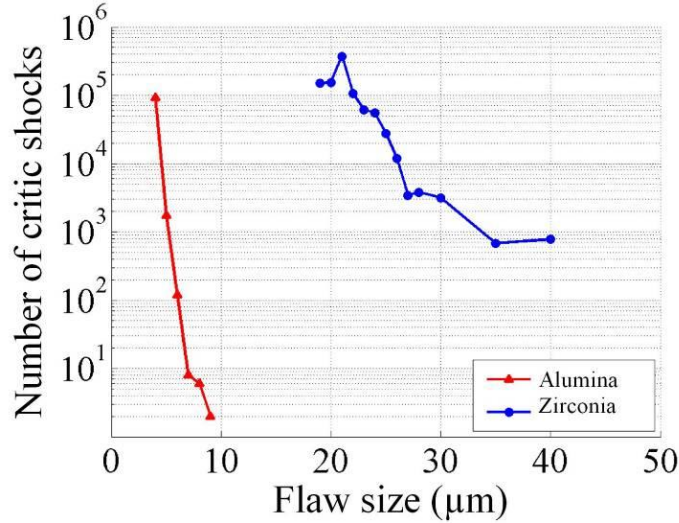

Figure 19.Influence of material on critical number of shocks

### 3. Comparison of maximum number of shocks

The main aim of this work is to present if zirconia or alumina materials involve the highest shock number according to the flaw size. If the shocks number is weak and the flaw size is lower than 10 µm, one might suggest that flaws, providing from the sintering process, could involve the structure failure under shocks.

Finally, four modellings were investigated: head diameters of 28mm and 32mm for alumina and zirconia. The inclination was constant and equal to 45°, the load was of 9kN and the microseparation was of 1.3mm. Alumina 32mm with a thick cup appears to be more resistant to shocks than alumina diameter 28mm, Figure 20a. However, the zirconia diameter does not exhibit a significant influence, Figure 20b. The critical flaw sizes of zirconia are three times higher than the ones of alumina.

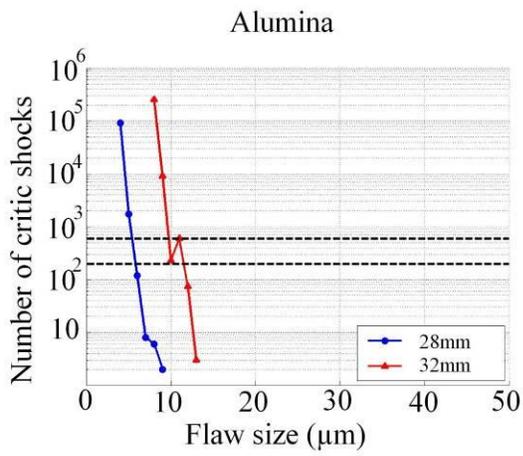 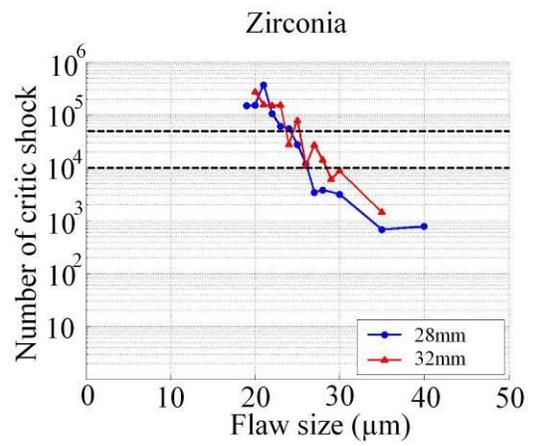

a) b)

Figure 20. Influence of geometry on critical number of shocks. Critical number of shocks as function of flaw size for a) alumina and b) zirconia. Dotted lines correspond to the interval of lifetime found experimentally.

## DISCUSSION

**Finite element analysis**

The finite element analysis allowed highlighting the significant parameters for the crack growth modelling (stresses) and establishing the mechanical behaviour description. Concerning the values of stresses, it is worth noting that the head is subjected to compressive stresses whereas the cup is mainly subjected to tensile stresses more harmful. In addition, a weak microseparation has a huge influence on the location of stresses. Thus, the microseparation produced a stresses concentration at the rim of the cup, the experimental tests exhibit too a fracture at the rim of the cup. Rebounds and conflicts between head and cup, when the hydraulic jack go down involved two wear zones, at the upper and the lower part of the cup, which matched with the two wear stripes during the experimental tests.

Moreover, the stress maximum only depended on the head velocity before impacting the cup. In fact, this maximum is also influenced by the mechanical behaviour of the assembly.

Three inclinations were then tested: 30°, 45° and 60°. When the cup is mounted at 60° the smaller maximum of stress is obtained regardless of the microseparation.

This simulation also highlighted the differences between the two types of tested bioceramics, zirconia and alumina. Alumina is submitted to higher compressive stresses than zirconia since its Young's modulus is two times higher. Finally, the worst case would be described in the following conditions: load of 9kN; microsepration of 1.3mm, inclination of 30° and alumina material. The inclination depends on the anatomical factors. The usual surgical inclination is within the range from 35° to 60° [23]. From Kummer et al 1999 [24], the hip rotation is not limited by the inclination. The inclination affects the stresses distribution on the cup. From the presented results, we should recommend that the inclination should be higher than 45°. However, we did not investigate all mobility angles, anteversion, varus and valgus prosthesis orientation and so on.

**Crack growth modelling**

Nevertheless, severe modelling hypotheses were considered, these results (in terms of lifetime and location of critical flaws) are in good agreement with the experimental tests, see Figure 16. The algorithm of crack growth, based on Todinov's equations, was used to estimate the probability of failure which depended on the stresses on the material and the rate growth of the flaws. The iterative nonlinear equation (Eq 5) used to evaluate the evolution of a flaw size was tested with several growth laws and its deterministic behaviour was satisfied. Thus, this deterministic equation was inserted in the algorithm for estimating ceramic elements lifetime.

The validation of results was made by comparing theoretical and experimental lifetime for zirconia-zirconia contact. The order of magnitude, i.e. the lifetime, was similar regardless of the fact that crack growth laws were taken from literature [15] and did not exactly correspond to tested biomaterials. Additionally, the theoretical location of critical flaws likely to induce a crack was in agreement with the optical observations, see Figure 16; flaws were located on the wear stripes on the femoral head surface. This comparison with experimental results, it got a clue for validating the model. A model combining FEA and post-processing was also implemented for crack growth simulation.

**Most significant parameters**

The studied parameters were: material, head diameter (consequently cup diameter), cup geometry applied force, inclination and microseparation. Excluding the applied force which has an obvious influence on the lifetime of the prosthetic elements, microseparation and inclination of the cup seem to be both significant parameters.

The inclination has an influence on the probability of failure. From figure 14, one might suggest that the inclination of 60° involved the lowest failure probability, whereas when inclined at 30° stresses are too much higher. Nonetheless, a very vertical inclination might increase the risk of impingement, which consists of intermittent contacts between the cone and the rim of the cup. This phenomenon has been reported to produce several in vivo fractures of ceramic cups [25]. One might expect that the implants positioning is a key factor for improving the implants lifetime. The positioning includes the inclination, as described in this work, but it involves the global positioning of the femoral stem, i.e. the location of the cup in the acetabulum, the height of head on the femoral neck, the sizing of the assemble head and cup, etc. All these factors should be relevant and they are not investigated during this work. It should be relevant for treating this problem about ceramic materials as artificial hip joint. This problem is related to the kind of head and cup assembly. Indeed, a smooth (polymer as UHMWPE)-hard (metal or ceramic) assembly involves no significant mismatches if the centres of the spheres are not coinciding. On the contrary, a hard-hard assembly involves severe degradations if mismatching. We got a clue to show that the shocks, due gait cycle with microseparation, influence hugely the lifetime of ceramic cup. These mismatchings should decrease the lifetime duration. Moreover, the clearance, the deviation of head and cup radii, plays a significant role [26].This radius difference is about few hundred micrometers. One might suggest for optimizing the clearance in order to improve the CoC assembly lifetime.

In the model, after the impact head-cup, only the compressive stresses on the head were taken into account. Nevertheless, there could be important tensile stresses on the femoral head created by the taper-fit. Further investigations should be made in order to confirm if these tensile stresses could lead to fracture on the femoral head as observed *in vivo*.

**Limitations of the investigating modelling**

The elastic recovery of the actuator of the cup was not taken into account. Additional investigations will be in progress for taking into account the mechanical impedance of the device. Concerning the FEA, the contact head-cup was supposed frictionless, whereas during the tests, a wear stripe due to sliding movement was observed. The materials were assumed perfectly elastic, but at microscale some plastic deformations could occur. A limited number of elements were chosen to have reliable results, only for the more solicited zones. Thus, it would be interesting to refine the mesh in the volume, with the aim of confirming that stresses are mostly located on the surface. Nevertheless, refining the mesh would increase the calculation time. In this study, the purpose was to obtain reliable results even by meshing roughly, since 2D results were interpolated with a refined mesh which multiplied the elements. Crack growth laws, crack rate as function of the stress intensity factor, were taken from literature. These values should be determined for the materials tested in the present study but it would need a lot of time. Therefore, the crack growth law was developed only for a flaw supposed spherical. In addition, theoretical estimations of lifetime are in agreement with the experimental results for zirconia and alumina, which suggests that the crack growth curves used are not too different to the actual case.

# CONCLUSIONS

The modelling confirmed the hypotheses concerning the assembly head-cup. It has been demonstrated that the minor microseparation involves a huge influence. Moreover, the cup is subjected to tensile stresses whereas the head is subjected to compressive stresses. The modellings also confirmed the location of high stressed zones at the rim of the cup, corresponding to location of wear stripes on the heads from the experimental tests. Crack growth algorithm is based on the stresses calculated by finite elements analysis. Despite the simplifying hypotheses and the crack growth curves taken from literature, results are on agreement with the experimental observations. In fact, all the simulated flaws likely to induce rupture are located on the cup surface corresponding to wear stripes on the head. Furthermore, the equation to calculate probabilities of failure takes into account the porosity. This model combining FEA with crack growth modellings confirmed the hypotheses made during experimental tests. In addition, this model served to theoretically testing parameters not studied experimentally. Thus, inclination of the cup seems to play a significant role in the hip prostheses degradation. It has also been shown that microseparation produces wear stripes and flaws on the cup surface which might lead to the fracture. Given that wear stripes appear early on the head during in vitro tests, more attention must be paid to this phenomenon even if it appears occasionally in vivo. Crack growth modellings revealed other significant parameters in cup lifetime like inclination of the cup and porosity of the bulk material. Thus, for an alumina cup with a porosity higher than 0.7%, the less critical flaw leads to a failure probability higher than 0.9. In fact, for porosity 0.8%, these flaw sizes lead in densities higher than those for zirconia, thus the minor critical flaw in alumina will lead probabilities of failure near to unity. Consequently, estimating lifetime by means of critical number of shocks is also valid for alumina.

An interesting development concerning FEA could be refining the cup mesh in 3D. One should pay attention on calculating more accurate stresses and verifying that they are mainly located on the surface rather than the bulk cup. A sub-model could be made with only the rim of the cup since the bottom does not take part in the global kinetics of the head-cup assembly. Given that at the microscale some plastic deformation could occur in the cup, plastic properties should be set to alumina and zirconia. A multiscale approach should be developed for understanding the crack initiation and evolution from the microscopic, which is grain scale to macroscopic scale. Concerning the crack growth algorithm, the three dimensional flaws should be investigated instead of only those on the surface and with a non spherical shape but ellipsoidal. Thus, it would be necessary to give a specific orientation to each flaw. This study could be the base for developing new tests as closer as possible to in vivo degradations. Nevertheless, it seems absolutely necessary to continue the shock tests since the principal cause of cup failure is the fracture instead of an excessive wear. Thus, if ceramics are shocks resistant they could be used without risk of fracture with a superior in vivo lifetime. Additional investigations are in progress for improving
Thus, this original study opens the way for further research.

**ACKNOWLEDGEMENTS**

The authors are grateful to:
- ANR (Agence Nationale de la Recherche) for granting the project 'Opt-Hip';
- Region Rhone-Alpes provided fundings for a PhD grant about this work;
- Nicolas Curt for his technical contribution about the shocks device.